\documentclass{aa}

\usepackage{mathptm}
\usepackage{graphicx}

\begin{document}

\thesaurus{09         
          (02.16.2;   
           09.19.2;   
           13.18.2)}  

\title{The supernova remnant RX J0852.0--4622: radio characteristics and
implications for SNR statistics}

\author{A.R.~Duncan\inst{1} \and D.A.~Green\inst{2}}

\offprints{A.R.~Duncan \\ (rduncan@mpifr--bonn.mpg.de)}

\institute{Max-Planck-Institut f\"{u}r Radioastronomie, Auf dem
           H\"{u}gel 69, D--53121 Bonn, Germany
\and
           Mullard Radio Astronomy Observatory,
           Cavendish Laboratory, Madingley Road, Cambridge CB3 0HE, U.K.
}

\date{Received 23 December 1999 / Accepted 31 August 2000}

\authorrunning{A.R.~Duncan \& D.A.~Green}
\titlerunning{The supernova remnant RX J0852.0--4622}

\maketitle

\begin{abstract}

We present new radio observations of the recently identified, young Galactic supernova
remnant (SNR) RX J0852.0--4622 (G266.2--01.2) made at 1.40~GHz with a resolution
of $14\farcm9$. These results, along with other radio observations from the
literature, are used to derive the extent, morphology and radio spectrum of the
remnant. The possible age and distance to this remnant are discussed, along
with the consequences of its properties -- especially its low radio surface
brightness -- for statistical studies of Galactic SNRs. The extended features
identified by Combi et al.~(\cite{combi}) are considered, and we conclude that
these are probably unrelated to the new remnant.  If RX J0852.0--4622 is
nearby, as is suggested by the available $\gamma$-ray data, then the range of
intrinsic radio luminosities for SNRs of the same diameter may be much larger
than was previously thought.

\keywords{Polarisation -- ISM: supernova remnants -- Radio continuum: general}
\end{abstract}

\section{Introduction}

Recently, Aschenbach (\cite{aschenbach}) reported the discovery of a young
supernova remnant (SNR) -- designated RX J0852.0--4622 -- from high-energy
X-ray data from the ROSAT All-Sky Survey. This new SNR appears near the
southeastern boundary of the Vela remnant (e.g.\ Milne~\cite{milne95};
Aschenbach et al.~\cite{aschenbach95}; Duncan et al.~\cite{duncan96}),
appearing in X-rays (with $E >$ 1.3~keV) as a nearly circular ``ring''
approximately $2\degr$ in angular diameter. Around the circumference of this
ring are a number of enhancements in the X-ray emission, the most prominent of
which appears near the northwestern perimeter. The available X-ray and
$\gamma$-ray data show the remnant to be comparatively young, with an age of
$\la 1500$~yr (Iyudin et al.~\cite{iyudin}; Aschenbach et
al.~\cite{aschenbach99}).

Following from this X-ray detection, Combi et al.~(\cite{combi}) reported a
radio detection of the SNR from the 2.42-GHz data of Duncan et
al.~(\cite{duncan95}). These authors present spatially filtered data from the
Parkes 2.42-GHz survey, along with results obtained from the 30-m Villa Elisa
telescope at 1.42~GHz (beamwidth $\sim 34\arcmin$).

The possibility of providing a more accurate age for this remnant was raised by
Burgess \& Zuber~(\cite{burgess}), who present a re-analysis of nitrate
abundance data from an Antarctic ice core. These authors find evidence for a
nearby Galactic SN $680\pm 20$ years ago, in addition to the known historical
supernovae (e.g.\ Clark \& Stephenson \cite{clark}), although it is not
possible to link this new SN definitively with RX~J0852.0--4622.

In this paper, we examine three sets of radio continuum data from the Parkes
telescope, at frequencies of 1.40, 2.42 and 4.85~GHz.  We use these data to
further investigate the radio structure of RX~J0852.0--4622.  Implications of
the radio characteristics of this remnant for statistical studies of SNRs are
then considered.

\section{Radio data}

The radio data presented here come from three principal sources, at frequencies
of 4.85, 2.42 and 1.40~GHz.  Characteristics of these data are given in
Table~\ref{table_data}.

First, 4.85-GHz data have been obtained from the Parkes-MIT-NRAO (PMN) survey
images (Griffith \& Wright~\cite{griffith}). These images were observed using
the 64-m Parkes radio telescope, and have an angular resolution of
approximately $5\arcmin$. Processing of the PMN observations has removed
large-scale information ($\ga 1\degr$) from the data.  Nevertheless, the PMN
images are a useful source of higher resolution information, and are often able
to trace structures of large angular size through associated smaller-scale
emission components (e.g.\ Duncan et al.~\cite{duncan97}).

Second, 2.42-GHz data surrounding RX~J0852.0--4622 have been observed as part
of a larger survey presented by Duncan et al.~(\cite{duncan95}).  These data
have a resolution of $10\farcm4$ and include linear polarisation information.
Some results from these data pertaining to the Vela region have been presented
by Duncan et al.~(\cite{duncan96}). These data were used by Combi et
al.~(\cite{combi}) to make the radio detection of RX~J0852.0--4622.

Third, 1.40-GHz observations of the region containing the remnant were obtained
in 1996 September, as part of a larger survey of the Vela region at this
frequency.  Some of these data have already been used by other authors (e.g.\
Sault et al.~\cite{sault}).

The observing procedure employed for these 1.40-GHz data was analogous to that
used for the 2.42-GHz survey (Duncan et al.~\cite{duncan95}). The telescope was
scanned over a regularly-spaced coordinate grid, at a rate of $6\degr$ per
minute, until the region of interest had been completely covered.  This
procedure was then repeated, scanning the telescope in the orthogonal
direction.  Stokes-$I$, $Q$ and $U$ data were recorded.  The source PKS
B1934--638 was used as the primary gain calibrator for the observations. The
flux density of this source was assumed to be 14.90~Jy at a frequency of
1.40~GHz.  The source 3C138 was also observed, in order to calibrate the
absolute polarisation position-angles.  The intrinsic polarisation
position-angle of 3C138 is $169\degr$ (Tabara \& Inoue~\cite{tabara}). After
the calculation and subtraction of appropriate ``baselevels'' from each scan,
each pair of orthogonally-scanned maps was combined.


\begin{table}
\caption[ ]{Details of the radio observations of RX~J0852.0--4622. Note that
the PMN survey data contain no information on scale-sizes of $\simeq 1\degr$
and larger. The rms noise is quoted per beam area.}\label{table_data}
\begin{tabular}{ccccc} \hline
\noalign{\smallskip}
 Frequency  &  rms noise  &    Angular     & Stokes    & Data origin     \\
 (/GHz)     &  (/mJy)     &  resolution    &           &                 \\
 1.40       &  20         &  $14\farcm9$   & $I, Q, U$ & this paper      \\
 2.42       &  17         &  $10\farcm4$   & $I, Q, U$ & 2.42-GHz survey \\
 4.85       &  8          &  $5\arcmin$    & $I$       & PMN survey      \\
\noalign{\smallskip}
\hline
\end{tabular}
\end{table}

\section{Location and morphology\label{section_location}}

The radio emission from RX~J0852.0--4622 is superposed upon a highly structured
region of the Vela remnant.  Much of this confusing emission is of similar
surface brightness to that seen from the new SNR. Furthermore, the very bright,
thermal region RCW~38 lies almost adjacent to the southeastern boundary of
RX~J0852.0--4622. The peak flux of RCW~38 is approximately 150~Jy~beam$^{-1}$
in the 2.42-GHz data.


\begin{figure*}
\centerline{\includegraphics[angle=270,width=12.0cm]{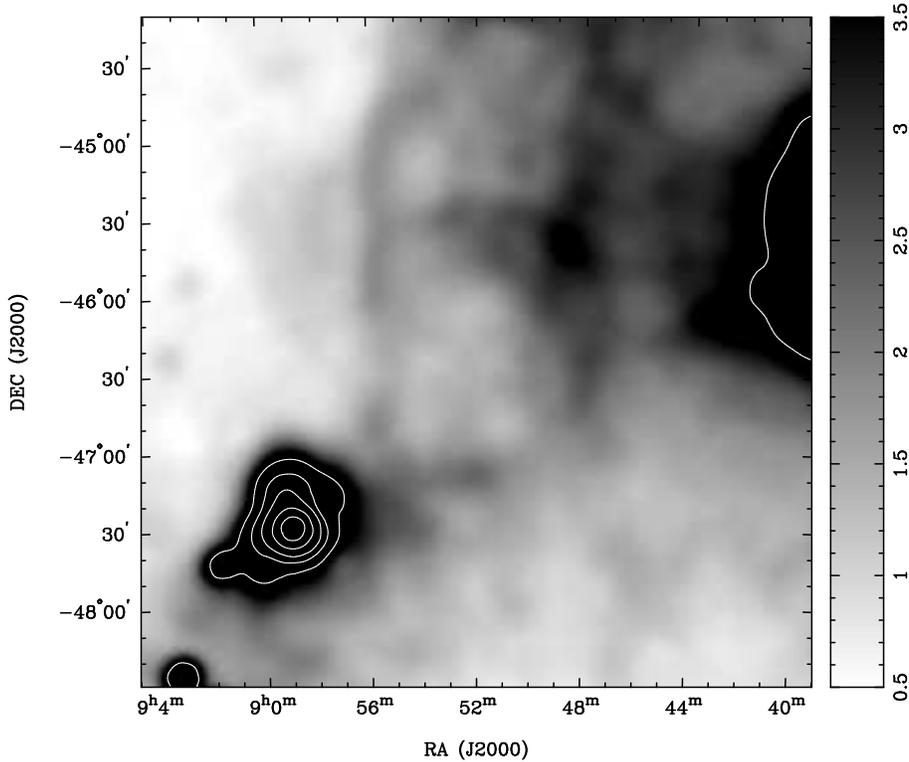}}
\caption[]{Unfiltered, Parkes 2.42-GHz survey image of RX~J0852.0--4622.
Although confused with larger structures, radio emission from the SNR --
approximately $1\fdg8$ in diameter in the centre of the image -- is clearly
visible. To the west, the brighter emission associated with Vela-X is visible.
To the southeast lies the bright \ion{H}{ii} region RCW~38. The angular
resolution is $10\farcm4$, and the rms noise is approximately
17~mJy~beam$^{-1}$.  The grey-scale wedge is labelled in units of
Jy~beam$^{-1}$. Contour levels are: 5, 10, 20, 50 and
100~Jy~beam$^{-1}$.}\label{fig_tpSfull}
\end{figure*}


\begin{figure*}
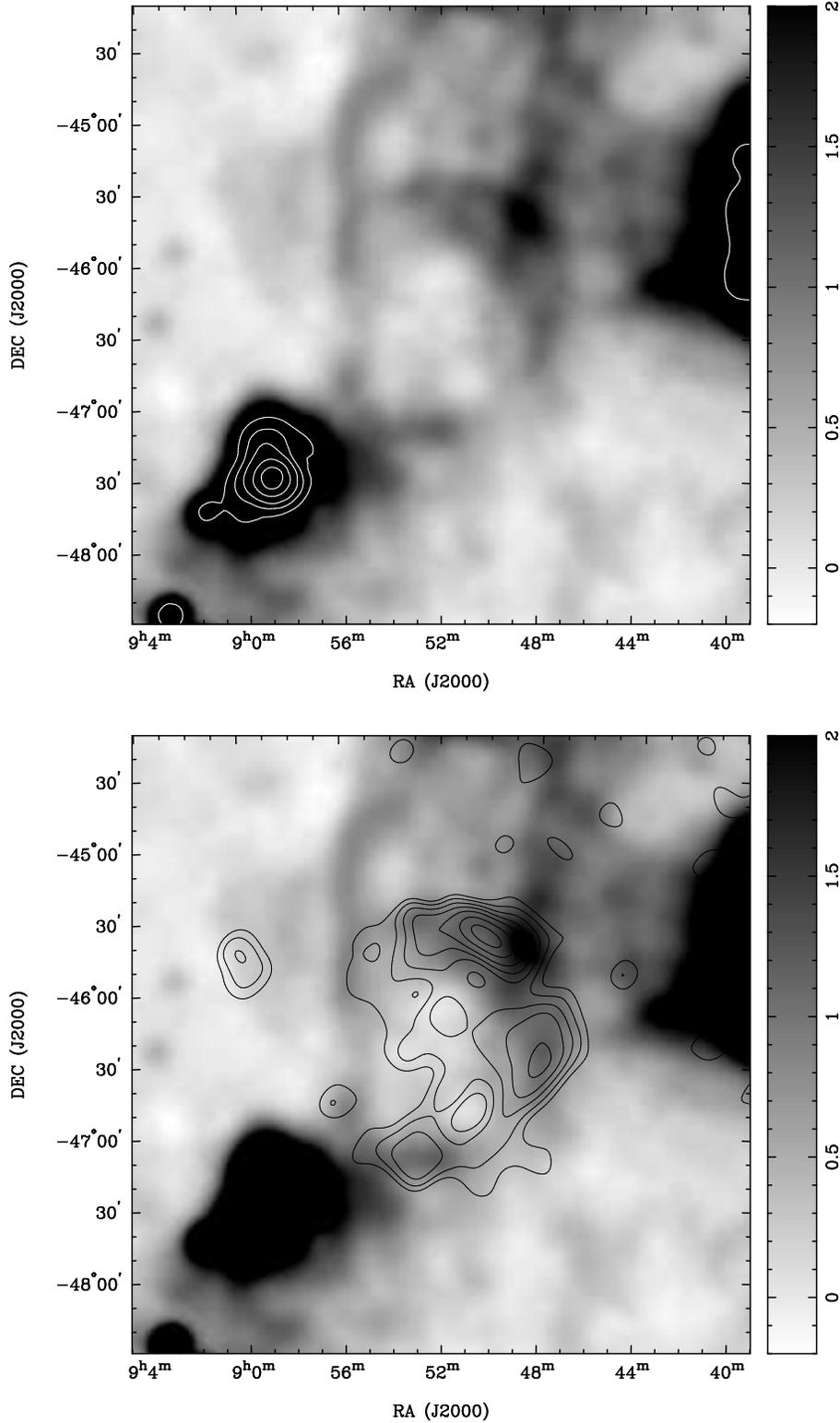

\centerline{\includegraphics[angle=270,width=12.0cm]{h1869.f2a}}
\vskip 5mm
\centerline{\includegraphics[angle=270,width=12.0cm]{h1869.f2b}}
\caption[]{The optimally-filtered, Parkes 2.42-GHz survey image of
RX~J0852.0--4622. A filtering beamwidth of $40\arcmin$ has been used, such that
larger-scale emission components have been eliminated. The SNR is now more
obvious, although it is still heavily confused with emission from the much
larger Vela remnant. The angular resolution is $10\farcm4$, and the rms noise
is approximately 17~mJy~beam$^{-1}$.  The grey-scale wedge is labelled in units
of Jy~beam$^{-1}$. The upper panel shows radio contours, with levels of: 5, 10,
20, 50 and 100~Jy~beam$^{-1}$. The lower panel shows contours of X-ray
intensity -- from the $12\arcmin$ resolution images of Snowden et
al.~(\protect\cite{snowden}) -- with levels of: 0.30, 0.40, 0.52, 0.65, 0.80, 1.0 and
$1.2 \times 10^{-3}$~counts~s$^{-1}$~arcmin$^{-2}$.}\label{fig_tpSfilt}
\end{figure*}


\begin{figure*}
\centerline{\includegraphics[angle=270,width=12.0cm]{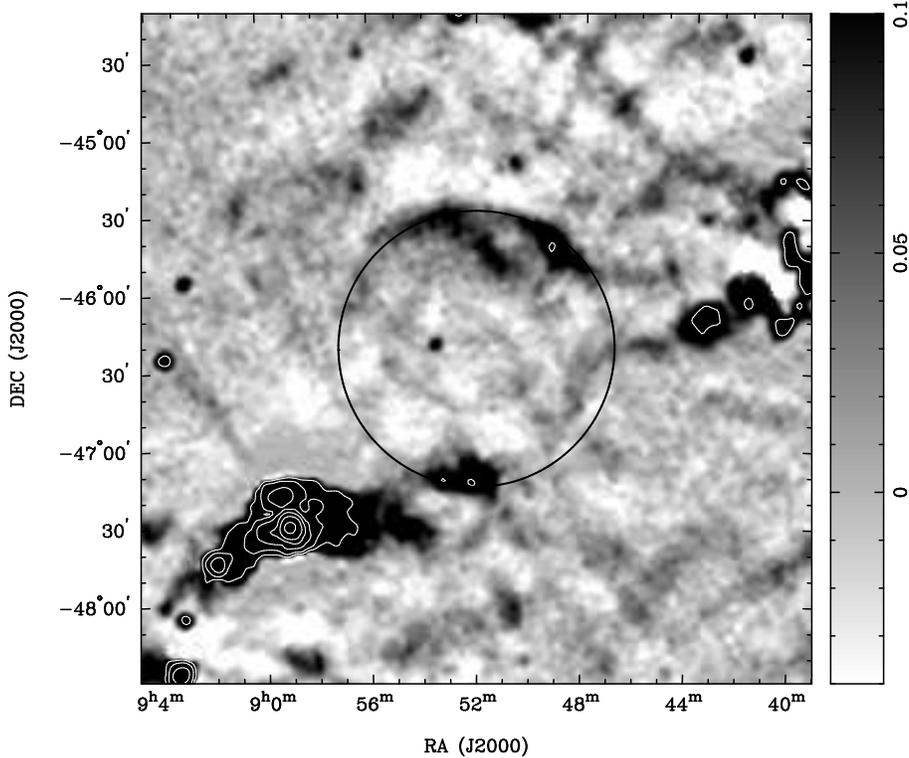}}
\caption[]{An image taken from the 4.85-GHz PMN survey, centred on the remnant
RX~J0852.0--4622. The angular resolution is $\simeq 5\arcmin$, and the rms
noise is approximately 8~mJy~beam$^{-1}$.  The grey-scale wedge is labelled in
units of Jy~beam$^{-1}$. The white contours denote intensities of: 0.2, 0.8, 3.0,
10, 30 and 80~Jy~beam$^{-1}$.  The black circle (see Sect.~\ref{subsection_485GHz})
is centred on the X-ray coordinates of the source and is $1\fdg8$ in angular
diameter.}\label{fig_pmn}
\end{figure*}

The presence of this confusing radio structure, both thermal and non-thermal,
meant that RX~J0852.0--4622 was not recognised as an SNR from pre-existing
radio observations of the region. Prior to the X-ray discovery of
RX~J0852.0--4622 the non-thermal emission in this region was thought to
emanate from the Vela SNR.

\bigskip
\subsection{The 2.42-GHz morphology\label{subsection_24GHz}}

The filtered 2.42-GHz image presented by Combi et al.~(\cite{combi}) clearly
shows the SNR to have a shell-like radio morphology.  This is even apparent in
unfiltered maps of the region, such as that presented in
Fig.~\ref{fig_tpSfull}. Indeed, the emission now known to be associated with
RX~J0852.0--4622 can be recognised in the radio images presented by Duncan et
al.~(\cite{duncan95}, \cite{duncan96}).  Combi et al.~(\cite{combi}) also
identify several additional features within their radio image, designated ``A''
through ``D'', which they suggest may represent extensions to the radio shell.
These will be considered in more detail in Sect.~\ref{subsection_extensions}.
It should be noted that -- possibly as a result of their filtering procedure --
the 2.42-GHz image presented by Combi et al.~(\cite{combi}) does not show
either the \ion{H}{ii} region RCW~38, or the bright, non-thermal emission from
Vela-X to the west.

Fig.~\ref{fig_tpSfilt} shows a spatially-filtered image of the region
surrounding RX~J0852.0--4622.  This image has been filtered using the ``bgf''
algorithm (e.g.\ Sofue \& Reich~\cite{sofue}), implemented within the NOD2
software package.  A number of filtering resolutions were used, and it was
(qualitatively) determined that the emission from RX~J0852.0--4622 was
optimally enhanced with a filtering resolution of approximately $30\arcmin$ to
$40\arcmin$ (in agreement with Combi et al.~\cite{combi}).

A filtering resolution of $40\arcmin$ was used for the radio data presented in
Fig.~\ref{fig_tpSfilt}.  This figure shows both the emission from
RX~J0852.0--4622 and the confusing structure more clearly.  Comparing
Fig.~\ref{fig_tpSfilt} with the unfiltered data presented in Fig.~\ref{fig_tpSfull},
it can be seen that removal of the large-scale structure does not have a major
effect on the appearance of the field.

The radio image of
the new remnant is dominated by two opposing arcs.  Some fainter radio emission
is visible on the remnant's western side, although there is no obvious
counterpart to the east. The brightest section of the radio shell lies to the
northwest, and appears approximately coincident with the brightest region of
the X-ray image. Comparing the radio with the X-ray emission
(Fig.~\ref{fig_tpSfilt}, lower panel), we see that the distributions of both
are generally similar, at least in as much as can be discerned from the
cluttered radio field.

\bigskip
\subsection{The 4.85-GHz PMN data\label{subsection_485GHz}}

Fig.~\ref{fig_pmn} shows data from the 4.85-GHz PMN survey from the same region
as shown in the previous figures. Although this survey is not optimised for
extended sources, the northern and southern sections of the limb-brightened
shell stand out clearly.

The black circle near the centre of Fig.~\ref{fig_pmn} fits the outer boundary
of the radio emission from both
Figs~\ref{fig_tpSfilt} and \ref{fig_pmn} well, and represents what we take to
be the outer boundary of the radio emission from RX~J0852.0--46221. This
boundary is $1\fdg8 \pm 0\fdg2$ in angular diameter, and is centred on the
X-ray centre of the SNR (as given by Aschenbach~\cite{aschenbach}).  Within the
uncertainties, this diameter is in agreement with that estimated by Combi et
al.~(\cite{combi}), who quote a value of $2\fdg1$, based upon the (lower
resolution) Parkes 2.42-GHz data alone.

Both the radio and X-ray data are consistent with a remnant centred on
Galactic longitude $266\fdg2$, latitude $-1\fdg2$.  Thus, we suggest a Galactic
designation of G266.2--01.2 for this SNR.

\bigskip
\subsection{Confusing structure within the field\label{subsection_confusion}}

As can be seen from Fig.~\ref{fig_tpSfilt}, a good deal of additional radio
structure is visible in the vicinity of RX~J0852.0--4622.  Over the remnant
itself, most of this structure takes the form of two diffuse ``filaments'',
each of which is $\simeq 20\arcmin$ wide.  These filaments are oriented
approximately north--south over the new SNR.  Beyond the northern boundary of
the remnant, the filaments begin to curve towards the west.

From larger images of the Vela region, such as have been presented by Duncan et
al.~(\cite{duncan95}, \cite{duncan96}), these filaments are known to curve
around Vela-X, forming almost a full quadrant of a circle.  The eastern arc (as
seen in Fig.~\ref{fig_tpSfilt}) is highly polarised (e.g.\ Duncan et
al.~\cite{duncan96}), and appears to represent the current boundary of the
shock from the Vela supernova event.

Interestingly, the confusing filaments from Fig.~\ref{fig_tpSfilt} are almost
completely absent from the PMN data.  This is because of the observing and data
processing procedures used as part of the PMN survey, coupled with the fact
that the confusing filaments lie approximately parallel to the scanning
direction of the telescope over the region of sky containing RX~J0852.0--4622.

\bigskip
\subsection{Extensions to the radio shell?\label{subsection_extensions}}

As mentioned in Sect.~\ref{subsection_24GHz}, Combi et al.~(\cite{combi})
identify a number of additional features within their radio image.  These
features were designated ``A'' through ``D'' in Fig.~1 of their paper, and
apparently extend for relatively large angular distances beyond the edge of the
RX~J0852.0--4622 shell (up to almost twice the radius of the remnant). Combi et
al.~(\cite{combi}) argue that these features may represent extensions to the
radio shell of RX~J0852.0--4622 (c.f.\ Aschenbach et al.~\cite{aschenbach95}).
It is of interest to consider these in more detail.

Examining the 2.42-GHz radio image of Combi et al.~(\cite{combi}), we find that
radio features ``A'' and ``C'' appear to be sections of the much more extensive
``arc'' structures discussed in Sect.~\ref{subsection_confusion}. These arcs
can be traced in Fig.~\ref{fig_tpSfilt} for several degrees, up to the northern
edge of the figure (i.e., beyond the boundary of RX~J0852.0--4622). Larger radio
images of the region show that these features continue for many degrees
further, in both total-power and polarised intensity. Feature ``B'' appears to
be an isolated, slightly extended source with no obvious connection to the new
remnant (even in the unfiltered image presented in Fig.~\ref{fig_tpSfull}).
Finally, feature ``D'' corresponds to the X-ray feature ``D/D$'$'' as
identified by Aschenbach et al.~(\cite{aschenbach95}).
Aschenbach~(\cite{aschenbach}) notes that this feature is also a source of hard
X-rays, but that this emission is associated with a much lower temperature
spectrum than that from RX~J0852.0--4622.

We suggest, therefore, that none of the possible ``extensions'' identified by
Combi et al.~(\cite{combi}) are associated with RX~J0852.0--4622.

A further argument against an association between these features and the new
remnant is that the boundary of the RX~J0852.0--4622 shock is well fitted (in
both the PMN and Parkes 2.42-GHz survey data) by a circle.  This is consistent
with the radio morphologies of other young shell SNRs, such as Kepler
(Dickel et al.~\cite{dickel88}), Tycho (Dickel et al.~\cite{dickel91}), and the
remnant of SN1006 (Reynolds \& Gilmore~\cite{reynolds}), although we caution
that RX~J0852.0--4622 is considerably fainter than these and other young remnants.
We also note that the higher resolution PMN image (Fig.~\ref{fig_pmn}), although not
optimised for extended emission, shows no evidence for any connections between
the features noted by Combi et al.~(\cite{combi}) and the shell of the new remnant.

\bigskip
\subsection{The quasi-central source\label{subsection_source}}

The PMN image shows that a point-like (i.e.\ unresolved at a resolution of
$\simeq 5\arcmin$) source lies approximately $15\farcm5$ east of the apparent
centre of the remnant. This source is not coincident with either of the two
compact X-ray sources near the centre of the remnant that are discussed by
Aschenbach et al.~(\cite{aschenbach99}).  The PMN survey source catalogue
(Wright et al.~\cite{wright}) lists this source as PMN~J0853--4620, with a
4.85-GHz flux of $136\pm 11$~mJy. Being relatively faint, our ability to detect
PMN~J0853--4620 in the 2.42-GHz data is compromised somewhat by confusion,
coupled with beam dilution. Nevertheless, we can establish the 2.42-GHz flux to
be $\la 200$~mJy. This leads to a spectral index for this source of $\alpha \ga
-0.7$ (with $S \propto \nu^{\alpha}$).

The radio spectral indices of pulsar emissions are generally much steeper than
the $\ga -0.7$ estimated above for the source (e.g.\ Taylor et
al.~\cite{taylor}). Furthermore, a flux of $136\pm 11$~mJy at a frequency of
4.85~GHz would be exceptionally high for a pulsar.  It is much more likely,
then, that the PMN~J0853--4620 source is extragalactic in origin, rather than
associated with RX~J0852.0--4622.

\bigskip
\subsection{Radio spectrum\label{subsection_spectrum}}

The presence of the confusing structure noted in
Sect.~\ref{subsection_confusion} makes accurate estimates of the integrated
remnant flux density difficult. The values for the flux density of the SNR
given below were estimated by integrating the emission within the boundary of
RX~J0852.0--4622, as defined by the circle seen in Fig.~\ref{fig_pmn} (lower
panel).  The integrated area extended approximately one beamwidth beyond this
circle, in order to include all the flux from the shell. The base level was
determined from flux minima near the centre of the remnant, as well as beyond
the eastern and southwestern edges of the shell. Fluxes contributed by the
confusing ``filaments'' seen to the western and eastern sides of the remnant
(as discussed in Sect.~\ref{subsection_confusion}) were
estimated and subtracted from the total, integrated flux.  Note that the
uncertainties in the integrated flux values are dominated by baselevel
uncertainty, rather than by uncertainties in the flux estimates of the
confusing structure.

We estimate the integrated fluxes of RX~J0852.0--4622 at 2.42 and 1.40~GHz to
be $33\pm 6$~Jy and $40\pm 10$~Jy, respectively. These values lead to a very
uncertain estimate of the spectral index, with $\alpha = -0.4\pm 0.5$
($S\propto \nu^{\alpha}$).  To better establish the spectral index of the
remnant, the method of ``T--T'' plots was used (e.g.\ Turtle et
al.~\cite{turtle}).

Estimates of the remnant spectral index were made from both filtered and
unfiltered images, using the T--T plot technique.  The northern section of the
shell was found to have a consistent, non-thermal index of $-0.40\pm 0.15$. The
southern section of emission exhibited a much flatter spectrum of $-0.1\pm
0.1$.  We believe this latter value to be unreliable, due to the proximity of
the southern section of the shell to the bright \ion{H}{ii} region RCW~38 and
its associated emission.  At the lower angular resolution of the 1.40-GHz data
(to which the 2.42-GHz data are also smoothed, for the purposes of spectral
index calculation), some of this thermal emission becomes confused with the
southern arc of RX~J0852.0--4622. We suggest that the value determined for the
northern shell section better represents the new remnant's radio emission.

Extrapolating the measured integrated flux to a frequency of 1~GHz (using a
spectral index of $-0.40\pm 0.15$), we determine a value of $47\pm 12$~Jy,
leading to an average surface brightness at this latter frequency of $(6.1\pm
1.5)\times 10^{-22}$~W~m$^{-2}$~Hz$^{-1}$~sr$^{-1}$. The above values are
summarised in Table~\ref{table_values}.

\bigskip
\section{Polarimetric observations\label{section_pol}}


\begin{figure}
\centerline{\includegraphics[angle=270,width=8.5cm]{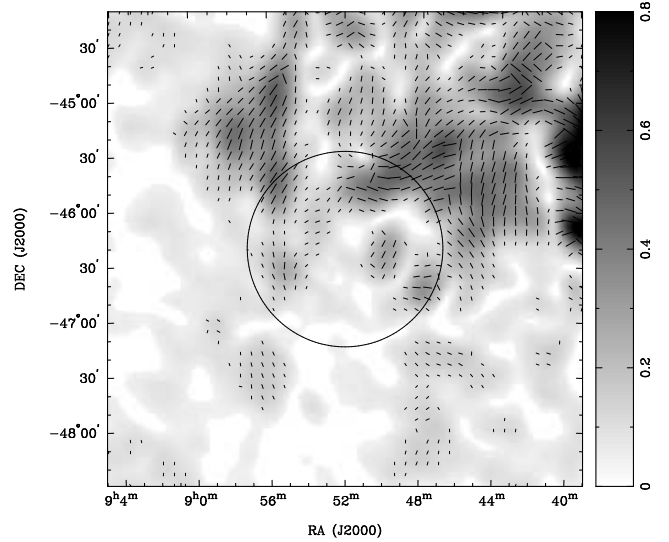}}
\caption[]{Grey-scale image showing the polarised intensities across the field
at 2.42~GHz, with a resolution of $14\farcm9$.  The 2.42-GHz Stokes-$Q$ and $U$
data have rms variations of $\simeq 23$~mJy~beam$^{-1}$ at this resolution. The
grey-scale image is blanked wherever the polarised intensity falls below
45~mJy~beam$^{-1}$. Orientations of the tangential components of the magnetic
fields are also shown.  These angles have been calculated from the 1.40- and
2.42-GHz data, assuming that the vector angles vary linearly with $\lambda^2$.
Errors in the derived angles are generally $\la 10\degr$. A vector is plotted
every $6\arcmin$, wherever both the 1.40- and 2.42-GHz polarised intensities
(at $14\farcm9$ resolution) exceed 0.1~Jy~beam$^{-1}$.  The grey-scale wedge
is labelled in units of Jy~beam$^{-1}$.}\label{fig_B}
\end{figure}


\begin{figure}
\centerline{\includegraphics[angle=270,width=8.5cm]{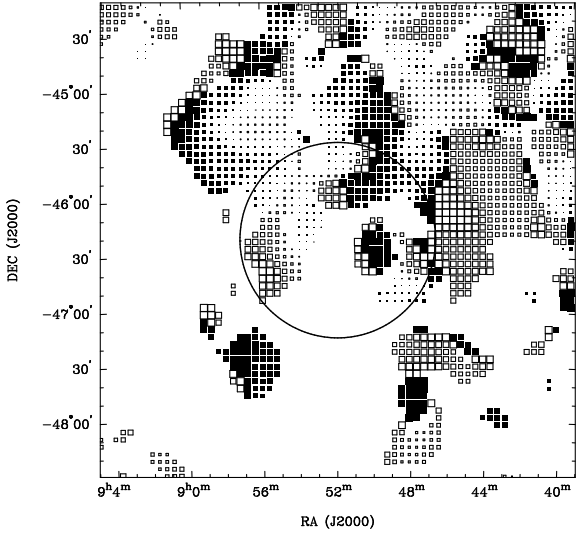}}
\caption[]{The derived rotation measure (RM) from the 2.42- and 1.40-GHz data,
assuming that the vector angles vary linearly with $\lambda^2$.  Filled squares
represent positive RM values, while empty squares represent negative values.
The size of each square is proportional to the magnitude of the RM, with the
maximum size corresponding to $\simeq \pm 51$~rad~m$^{-2}$.  The errors in RM
values are $\la 5$~rad~m$^{-2}$.  A square is plotted every $4\arcmin$. Data
have been blanked wherever the 1.40- or 2.42-GHz polarised intensity (at
$14\farcm9$ resolution) falls below 0.1~Jy~beam$^{-1}$.}\label{fig_rm}
\end{figure}

We have examined the 1.40- and 2.42-GHz polarimetric data in the field
surrounding RX~J0852.0--4622.  These data have been used to estimate the
Faraday rotation measures (RMs) across the field, and to calculate the
polarisation position-angles at zero wavelength (thereby estimating the
orientations of the tangential components of the magnetic fields within the
regions of polarised emission).  Note that these estimates assume the
polarisation position-angles to vary linearly with $\lambda^2$.

The polarised intensities from the 2.42-GHz data are shown in Fig.~\ref{fig_B},
at the lower resolution of $14\farcm9$.  Superposed upon this grey-scale image
are the orientations of the tangential component of the magnetic field. The RMs
derived across the field are shown in Fig.~\ref{fig_rm}.

The large circle shows the boundary of the SNR RX~J0852.0--4622. Much of the
polarised emission visible within the circle does not appear to be associated
with the new SNR.  For example, the polarisation detected from the eastern half
of RX~J0852.0--4622 is clearly associated with the prominent, eastern arc of
total-power emission.  The only region of RX~J0852.0--4622 to potentially
exhibit polarised emission is the northern section of the shell.  If associated
with the new remnant, the northern arc appears polarised to a level of
approximately 20\% at 2.42~GHz.  Other regions of the SNR exhibit no polarised
emission (such as the southern section of the shell), or are confused with
polarised structure from Vela.

The magnetic field vectors on the north side of the shell appear jumbled, with
no clear orientation evident.  This is in contrast to the field
orientations in other young, shell-type supernova remnants, which are
predominantly radial; e.g.\ Tycho (Wood et al.~\cite{wood}), Kepler (Matsui et
al.~\cite{matsui}) and SN1006 (Reynolds \& Gilmore~\cite{reynolds}).

However, it is questionable whether the detected polarised emission in this
northern section of the shell is attributable to RX~J0852.0--4622.  The
appearance of much of this RM structure is similar to that seen from the Vela
emission, beyond the RX~J0852.0--4622 shell's northwestern perimeter.
Furthermore, there is no discontinuity in either the RM values or the magnetic
field orientations near the boundary of the shell.  We suggest, therefore, that
the polarised emission seen throughout Fig.~\ref{fig_B} originates entirely
from the Vela remnant.  We note that this interpretation is also consistent
with the lack of polarisation observed from the southern arc of the
RX~J0852.0--4622 shell.

The fractional polarisation of the northern arc must then be $< 20$\% at
2.42~GHz.  As noted above, we detect no polarisation in the vicinity of the
bright, southern section of the shell, to a ($5\sigma$) limit of approximately
5\% at 2.42~GHz. We note that these low fractional polarisations are not
inconsistent with the polarimetric properties of other young shells, which
exhibit fractional polarisations of $\la 15$\% when the emission is well
resolved.


\begin{table} \begin{center}
\caption[ ]{Characteristics of the radio emission from RX~J0852.0--4622,
determined from the Parkes data. The spectral index measured for the southern
limb of the SNR appears to be confused with nearby thermal emission. The
surface brightness is given in units of
W~m$^{-2}$~Hz$^{-1}$~sr$^{-1}$.}\label{table_values}
\begin{tabular}{ll}
\hline
\noalign{\smallskip}
 Angular diameter                          &  $1\fdg8 \pm 0\fdg2$           \\
 Integrated 1.40-GHz flux density          &  $40\pm 10$~Jy                 \\
 Integrated 2.42-GHz flux density          &  $33\pm 6$~Jy                  \\
 $\alpha_{\rm northern}$ (from T--T plots) &  $-0.40\pm 0.15$               \\
 $\alpha_{\rm southern}$ (from T--T plots) &  $-0.1\pm 0.1$(?)              \\
 Integrated flux density at 1~GHz          &  $47\pm 12$~Jy                 \\
 Surface brightness at 1~GHz               &  $(6.1\pm 1.5)\times 10^{-22}$ \\
\noalign{\smallskip}
\hline
\end{tabular}
\end{center}
\end{table}

\section{Discussion\label{section_discussion}}

\subsection{RX~J0852.0--4622: young or not so young?\label{subsection_young}}

Several observations of RX~J0852.0--4622 suggest that it is a young SNR.  Most
notably, both the X-ray temperatures derived by Aschenbach~(\cite{aschenbach})
and the $\gamma$-ray work of Iyudin et al.~(\cite{iyudin}) imply an age of $\la
1500$~yr. Further, the more recent discussion of the $\gamma$-ray observations
by Aschenbach et al.~(\cite{aschenbach99}) suggest the remnant is $\sim 700$
years old. This is in agreement with the approximately circular radio
appearance of the remnant; a characteristic exhibited by other young,
shell-type SNRs. Even the cautious upper distance limit of approximately 1~kpc
provided by Aschenbach~(\cite{aschenbach}), which leads to a linear diameter of
$\simeq 30$~pc, implies the age of this remnant cannot be more than a few
thousand years.

However, there are some radio properties of this remnant are difficult to
reconcile with those of other young SNRs.
\begin{itemize}
\item The radio shell is far from complete.  The radio image of
RX~J0852.0--4622 is composed primarily of two, opposing regions of emission.
This is in contrast to other young, shell-type SNRs, which exhibit essentially
complete radio shells.  We note a possible resemblance to the structure
of SN1006, however, which also shows opposing arcs of emission.
\item The radio emission from this new SNR is of relatively low surface
brightness.  The mean radio surface brightness of this new remnant at 1~GHz
(see Table~\ref{table_values}) is a factor of 5 lower than that of SN1006. This
is significant, because the SN1006 remnant has the lowest surface brightness of
all the young, shell-type SNRs in current catalogues (e.g.\ Green~\cite{green})
-- see the further discussion in Sect.~\ref{subsection_statistics}.
\item The radio spectral index of $-0.40\pm 0.15$ determined herein for
RX~J0852.0--4622 is flatter than those of other young shells, which have
indices of $\la -0.6$ (e.g.\ Green \cite{green88}).
\end{itemize}

It is possible that some of the unusual radio properties of RX~J0852.0--4622 may
result from the SNR expanding into a hot, low-density region of the interstellar
medium; Aschenbach~(\cite{aschenbach}) determines an upper limit to this density
of approximately 0.06~cm$^{-3}$. If so, this would emphasise the the role played
by environmental effects on the detectability of young radio remnants.
Alternatively, this SNR may be just beginning to ``turn on'' at radio
wavelengths, although this scenario may be difficult to reconcile with the very
incomplete radio shell.

Further insights into the unusual radio properties of this SNR must await more
detailed investigations of the characteristics of both the remnant and the
environment into which it is expanding.

\subsection{The distance to RX~J0852.0--4622\label{subsection_distance}}

Unfortunately, the distance to RX~J0852.0--4622 is highly uncertain.  The
X-ray data of Aschenbach~(\cite{aschenbach}) provide an upper limit of
approximately 1~kpc, based on the lack of absorption, but suggest that the
remnant's distance could be as small as 200~pc. This lower limit is based upon
a comparison of the new remnant's surface brightness with that of SN1006.
However, SN1006 is atypically faint for known, young, Galactic SNRs (this is
further discussed in Sect.~\ref{subsection_statistics}).

The $\gamma$-ray data discussed by Iyudin et al.~(\cite{iyudin}) and Aschenbach
et al.~(\cite{aschenbach99}) suggest an age of approximately 700~yr, from a
comparison of the observed $^{44}$Ti line flux with that expected from SN
models, which corresponds to a distance of approximately 200~pc. However, the
interpretation of the $\gamma$-ray detection requires the assumption of both
the supernova shock velocity and the $^{44}$Ti yield. Iyudin et
al.~(\cite{iyudin}) note that increases in either of these quantities will lead
to an increase in the derived distance of the remnant.

We have examined 21-cm \ion{H}{i} observations in the region of
RX~J0852.0--4622, from Kerr et al.~(\cite{kerr}), in an attempt to find any
correlating features. However, since the remnant lies in a complex region in
Vela, no associated features in \ion{H}{i} could be found.

The ice core data of Burgess \& Zuber~(\cite{burgess}) may be able to provide
an accurate age for RX~J0852.0--4622, but these data are not able to constrain
the distance to the SNR without further assumptions. Furthermore, as noted
above, it is not possible to definitively associate the additional nitrate
spike with this SNR. If we assume the age of $\simeq700$~yr determined by
Burgess \& Zuber~(\cite{burgess}) to be accurate, then assuming an upper limit
to the mean shock velocity of $10^4$~km~s$^{-1}$ places the remnant at a
distance of $\la 460$~pc, with a linear diameter of $\la 15$~pc.

A value of $10^4$~km~s$^{-1}$ was nominated as an upper limit to the shock
velocity by Aschenbach et al.~(\cite{aschenbach99}), based on their
analysis of the X-ray data.  Should the mean shock velocity exceed this
value, the remnant would lie at a distance of $\ga 460$~pc, with a
correspondingly larger linear diameter.

In summary, the distance is very poorly constrained by current observational
data. Unfortunately, since the remnant is faint, is not detected optically, and
is in a complex region, many direct techniques used to determine the distance
to SNRs are not applicable in this case (e.g.\ \ion{H}{i} absorption,
association with other \ion{H}{i} features or molecular clouds, or optical
studies). Nevertheless, the radio properties of this remnant, even with the
present uncertainty in its age and size, have some interesting implications for
statistical studies of Galactic SNRs, which are discussed in the next section.

\subsection{Statistical implications\label{subsection_statistics}}

\begin{figure*}
\centerline{\includegraphics[angle=0,width=11.0cm]{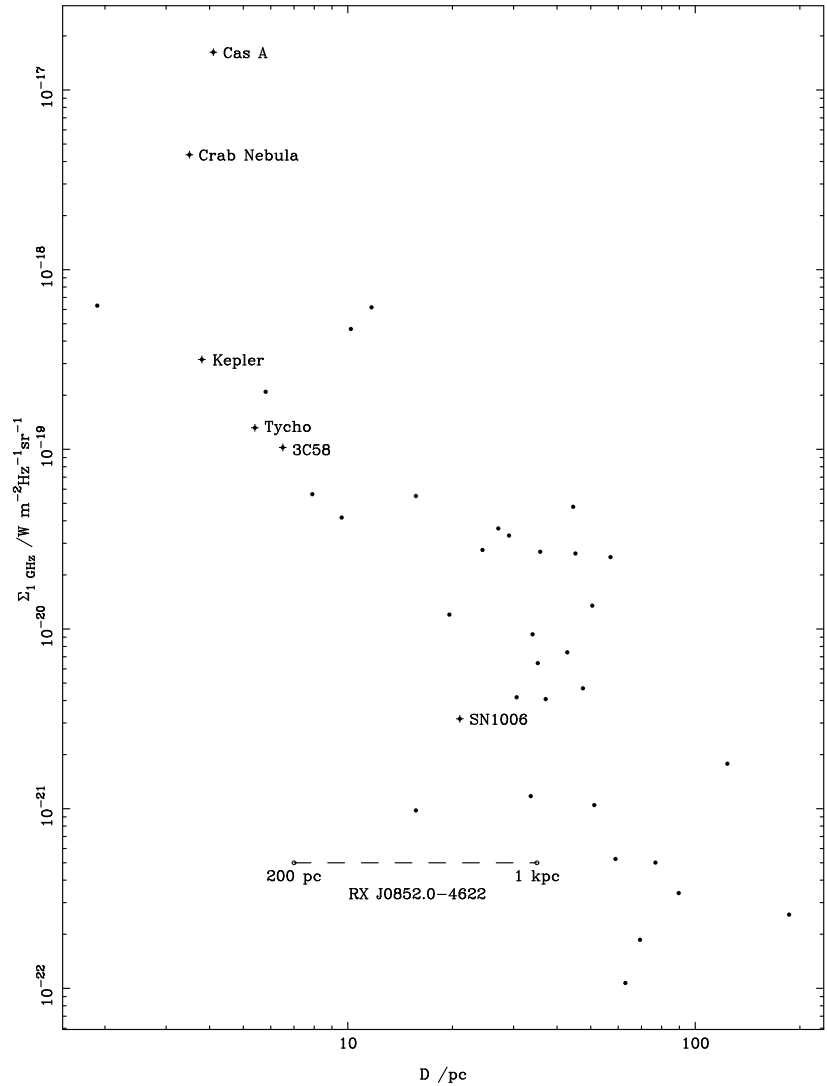}}
\caption[]{A surface-brightness vs.\ diameter ($\Sigma-D$) plot for Galactic
SNRs with relatively well known distance. RX~J0852.0--4622 is plotted for a
range of distances from 200~pc to 1~kpc, and known young SNRs are
labelled.}\label{fig_sigmadee}
\end{figure*}

Fig.~\ref{fig_sigmadee} shows a surface-brightness versus diameter plot for
Galactic SNRs for which there are reasonable distances available (Green
\cite{green84}, \cite{green91}, \cite{green}). RX J0852.0--4622 is plotted on
this figure with a range of diameters corresponding to distances from 200~pc to
1~kpc. The surface brightness of RX J0852.0--4622, of $\Sigma_{\rm 1~GHz} \sim
6 \times 10^{-22}$ W~m$^{-2}$~Hz$^{-1}$~sr$^{-1}$, is faint for a known
Galactic SNR -- among the faintest 20\% of catalogued remnants. This is less
than the nominal completeness limit of many radio surveys (e.g.\ Green
\cite{green91}). Note that although most faint remnants are thought to be old,
that the remnant of the SN of AD1006 is also faint, with $\Sigma_{\rm 1~GHz}
\sim 3 \times 10^{-21}$ W~m$^{-2}$~Hz$^{-1}$~sr$^{-1}$.
We also note that, whilst RX J0852.0--4622 is one of the fainter remnants
to appear on the $\Sigma-D$ plot, the only other remnant detected in
$^{44}$Ti $\gamma$-ray emission is Cas~A, which has the highest surface
brightness.


From Fig.~\ref{fig_sigmadee}, it is clear that the properties of RX
J0852.0--4622 are very unusual if it lies at the smaller distances suggested by
the $\gamma$-ray data. If the SNR is at a distance $<300$~pc, its diameter is
less than 10~pc, but its surface brightness is two or more orders of magnitude
less than other known young SNRs with similar diameters (e.g.\ Kepler's SN,
Tycho's SN, and 3C58). This would have important consequences for statistical
studies of Galactic SNRs (see Green \cite{green91}), as the range of $\Sigma$
-- or, equivalently, luminosity -- for a given $D$ may be even wider than was
previously thought. This in turn would imply that the observational selection
effect of faint SNRs being difficult to detect is important, not only for old
SNRs, but also for young SNRs. The low radio surface brightness of
RX~J0852.0--4622 indicates that a fraction of young SNRs may be faint at radio
wavelengths.  The available sample of young SNRs (i.e.\ historical events) is
small, however, so it is not possible to meaningfully estimate this proportion.

On the other hand, if the remnant is as distant as 1~kpc, then although it is
faint for it's diameter of $\sim 30$~pc, it is not strikingly unusual.

\bigskip
\section{Conclusions\label{section_conclusions}}

We have presented various radio observations of the newly recognised SNR
RX~J0852.0--4622, which clarify its size and morphology. Several features
possibly associated with the remnant by Combi et al.~(\cite{combi}) are
examined, and it is concluded that these are probably unrelated to
RX~J0852.0--4622.

Although the distance and age of RX~J0852.0--4622 are not well determined, its
faintness has some interesting implications for the statistical study of SNRs,
namely that the surface-brightness limits of current radio surveys may miss
faint, young remnants, as well as faint, old remnants. Clearly an accurate
distance determination for this SNR is important, although this is difficult,
given its faintness and the fact that it lies in a complex region of the sky,
confused with emission from the much larger Vela SNR.

\bigskip
\begin{acknowledgements}

The Australia Telescope is funded by the Commonwealth of Australia for
operation as a National Facility managed by CSIRO. ARD is an Alexander von
Humboldt research Fellow and thanks the Stiftung for their support.
The authors gratefully acknowledge P.\ Slane for helpful comments
and suggestions on the manuscript.

\end{acknowledgements}

\end{document}